\documentstyle[aps,epsf,prl,floats]{revtex}

\def\sb#1{$_{#1}$}

\begin{document}

\draft

\wideabs{
\title{Microscopic Charge Inhomogeneities in Underdoped
La$\bf _{2-x}$Sr$\bf _x$CuO$\bf _4$: Local Structural Evidence}

\author{S. J. L. Billinge, E. S. Bo\v zin and M. Gutmann }
\address{Department of Physics and Astronomy and Center for
Fundamental Materials Research, Michigan State University, 
East Lansing, MI 48824-1116.}
\author{
H. Takagi}
\address{Institute for Solid State Physics, University of Tokyo,
7-22-1, Roppongi, Minato-ku, Tokyo 106, Japan}

\date{\today}

\maketitle

\begin{abstract}
We present local structural evidence for the existence of charge
inhomogeneities at low temperature in underdoped and optimally doped
La$_{2-x}$Sr$_x$CuO$_4$.  The inhomogeneities disappear for $x\ge 0.2$.
The evidence for the charge inhomogeneities comes from an anomalous
increase in the in-plane Cu-O bond length distribution in the
underdoped samples as well as evidence for CuO\sb{6} octahedral tilt
inhomogeneities in the intermediate range structure. Preliminary
analysis of the temperature dependence of this phenomenon indicates
that the inhomogeneities set in at temperatures in the range 
$60 K<T_{co}<130~K$ which depends on doping.

\end{abstract}
}

\section{introduction}

One of the most interesting and intensively debated subjects in the
field of high-temperature superconductivity is the possibility that
the charge distribution in the electronically active CuO\sb{2} planes
of the cuprates is inhomogeneous.  Charge inhomogeneities, in the form
of ``stripes'' of charge in an insulating background, have been
observed in closely related insulating materials such as layered
nickelates (La$_{2-x}$A$_x$NiO$_{4+\delta}$,
A=Sr,Ba)~\cite{tranq;prl94,sacha;prb95}, perovskite
manganites\cite{chen;prl96,mori;n98} and neodymium co-doped
La$_{2-x}$Sr$_x$CuO$_4$~\cite{tranq;n95,tranq;prb96}. In this latter
system, the charge stripes are also seen in weakly superconducting
samples\cite{tranq;prl97i}. However, the phenomenology suggests that
the charge-stripes compete with superconductivity since the most
stable stripes (highest charge-ordering temperature) coincide with the
lowest superconducting transition
temperatures~\cite{tranq;prl97i,ichik;cm99}.  So what is the great
interest in this phenomenon for understanding high-temperature
superconductivity?

The interest comes from three directions.  First, from a theoretical point
of view it is apparent from a number of studies of strongly correlated
electron models that an instability towards charge phase separation is
an intrinsic property of these systems%
~\cite{zaane;prb89,schul;prl89,emery;pc93,white;prl98,marti;cm00}. 
It is clearly important to explore
the implications of this profound observation.  Secondly, despite great effort
over a large number of years there is no single theory that adequately explains
all of the phenomenology of the high-temperature superconductors.  Perhaps
a theory which has as an underlying principle an inhomogeneous charge 
distribution will be more successful.  A number of candidates have emerged
in recent years%
~\cite{emery;pnas99,bianc;ssc97,caste;cm00,gorko;js99,ovchi;prb99,hasse;prl99,%
goode;prb94,phill;pmb99,marki;jpcs97}.  Finally, as a further
motivation, there are a number of experimental observations which are
rather naturally interpreted in terms of short-range ordered
fluctuating (or quasi-static) charge inhomogeneities in
superconducting
systems~\cite{ichik;cm99,mason;prl92,thurs;prb92,yamad;prb98,mook;n00,hunt;prl99}.
It is clearly necessary to establish beyond doubt whether fluctuating
charge stripes exist in the superconducting cuprates and also to
characterize their presence as a function of temperature and doping.

\begin{figure}[tb]
\begin{center}$\,$
\epsfxsize=3.2in
\epsfbox{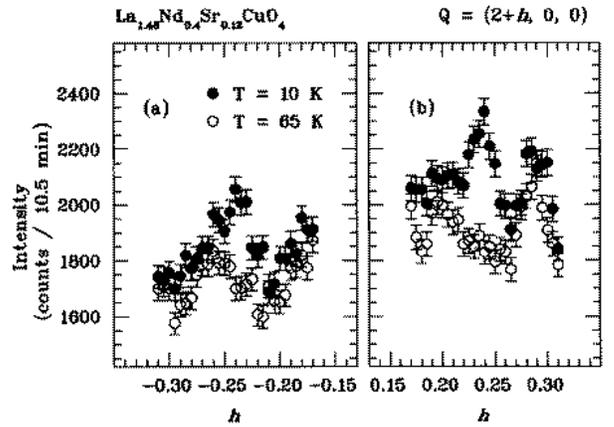}
\end{center}
\caption{Neutron nuclear superlattice peaks from 
La$_{1.475}$Nd\sb{0.4}Sr$_{0.125}$CuO$_4$.  This figure is
adapted from Tranquada {\it et al.}~\protect\cite{tranq;prb96} and shows
their data.
}
\protect\label{fig;tranqdata}
\end{figure}

What is the most convincing evidence for the presence of static
charge-stripes in the cuprates?  The seminal result that really
changed the way that people think about these materials was the
observation of neutron nuclear superlattice peaks in
La$_{1.475}$Nd\sb{0.4}Sr$_{0.125}$CuO$_4$ by Tranquada {\it et
al.}~\cite{tranq;n95}.  The data themselves are reproduced in
Fig.~\ref{fig;tranqdata}~\cite{tranq;prb96}.
The peaks are small and
it took many years (similar Nd doped samples were first studied by
Crawford {\it et al.}~\cite{crawf;prb91} and 
B\" uchner {\it et al.}~\cite{buchn;pc91} as early as 1991), and some clear
insight, to find them.  However, the argument that they originate from
charge ordering is compelling.

What do the superlattice peaks tell us?  First, their $Q$-dependence
($Q$ is the momentum transfer of the scattered neutron) indicate they
are nuclear, and not magnetic peaks. Their periodicity and temperature
dependence, especially in relation to magnetic superlattice peaks seen
in the same material, suggest that they are charge ordering peaks
similar to (but different from) those known in the
nickelates~\cite{tranq;prl94,sacha;prb95}.  The observation of sharp
peaks tells us that the charges are fairly long-range ordered in these
samples and that they are static or quasi-static.  The position in
reciprocal space of the peaks suggests the charge-order to be striped.
Finally, and most importantly from the point of view of this paper,
the {\it peaks originate from a structural distortion}.  We know this
because the neutrons being scattered are uncharged objects and don't
couple to the charges (there is an interaction with the electrons
through the spin-spin interaction which gives rise to magnetic
scattering; however, it has been established that these charge order
peaks are nuclear peaks).  The inference is that charges order into
static stripes.  These stripes locally distort the lattice giving rise
to a structural modulation which results in neutron nuclear
superlattice peaks.  This tells us, beyond doubt, that the static
stripes are coupled to the lattice.

What don't the observed superlattice peaks tell us?  The superlattice
peaks don't tell us much about the nature of the structural distortion
which gives rise to them.  Basic diffraction theory teaches us that the
existence and position of Bragg peaks gives basic information about
the size and shape of the periodic unit cell.
Information about the positions of atoms within the unit cell (and
changes in these positions when the charge-stripes form) is contained
in the crystallographic structure factor, which gives the relative
intensities of all the Bragg peaks.  To understand the nature of the
distortion, it would be necessary to measure accurately the 
intensities of a relatively large number of the superlattice (and
principal) Bragg peaks.
The nature of the structural distortion is of interest because it yields
microscopic information about the nature of the electron-lattice
interaction giving rise to the structural distortion.

We would like to address two outstanding questions: First, is there
any {\it structural} evidence for locally fluctuating charge-stripes
in superconducting samples in the absence of charge-order superlattice
peaks?  Second, can we determine the nature of the structural
distortion induced by the static (or slowly moving) charge-stripes?
Answering the first question requires us to determine some kind of
structural order parameter for the existence of charge
inhomogeneities: the local structural equivalent of the charge order
superlattice peak.  Once we have found this, it is clearly important
to establish the universality of the behavior among all high-T\sb{c}
materials if its importance to the phenomenon is to be established.

\begin{figure}[tb]
\begin{center}$\,$
\epsfxsize=2.8in
\epsfbox{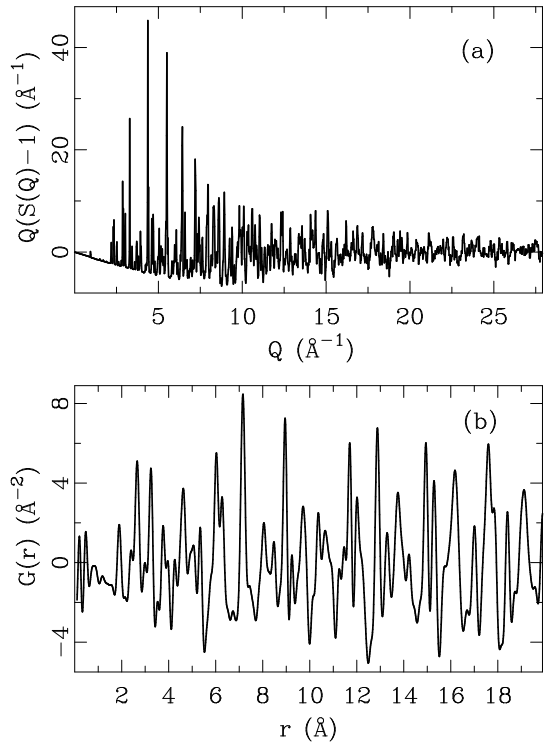}
\end{center}
\caption{A representative data-set from La$_{2-x}$Sr$_x$CuO$_4$.
(a) The total scattering structure function expressed in the form
that is Fourier transformed to obtain the PDF. (b) The reduced PDF,
$G(r)$, which is the direct sine-Fourier transform of the data in (a).
Data were collected on SEPD at the IPNS at Argonne National Laboratory.
}
\protect\label{fig;egdata}
\end{figure}

Since, in general, the charge (and therefore lattice) inhomogeneities
are not long-range ordered it is necessary to use a local structural
probe.  We have used the atomic pair distribution function (PDF)
analysis of powder neutron diffraction data.  In this technique
neutron powder diffraction data are measured with high accuracy over a
wide range of $Q$ using a pulsed-neutron source. The data are
corrected for experimental effects such as detector efficiency,
absorption, multiple scattering and so on.  The data are then
normalized with respect to the incident flux and number of scatterers,
respectively, to obtain the single-scattering total scattering
structure function, $S(Q)$.  An example of a structure function and the
resulting PDF are shown in Fig.~\ref{fig;egdata}.
The data reduction process is quite
standard and well
controlled~\cite{egami;b;utbp00,wagne;jncs78,billi;prb93} resulting in
highly accurate atomic pair distribution functions.  
The data were analyzed using the PDFgetN 
program~\cite{peter;jac00}.  The PDF provides
a measure of the probability of finding a neighboring atom at a
distance~$r$ from another atom.  As with conventional powder
diffraction, three dimensional structures can be inferred from these
one dimensional functions by regression modeling techniques.  When
applied to well ordered crystalline materials, results in quantitative
agreement with Rietveld refinement are obtained~\cite{proff;prb99,gutma;prb99}
verifying the accuracy of the technique.  Because the {\it total
scattering}, including Bragg and diffuse intensity, is Fourier
transformed to obtain the PDF, local structural distortions away from the
average structure are also obtained.  This has been shown convincingly
in recent PDF studies of In\sb{1-x}Ga\sb{x}As semiconductor alloys
where the average structure (the ``virtual crystal'' structure)
predicts a single In/Ga-As bond length but the PDF clearly resolves a
shorter Ga-As and longer In-As bond present in the local 
structure~\cite{petko;prl99} in
agreement with earlier XAFS~\cite{mikke;prl82}
results.  The PDF accurately measures the {\it local}
and {\it intermediate} range structure on the nanometer 
length-scale.
Since it is a neutron powder diffraction measurement, it is also a 
{\it bulk} measurement.

\section{ Search for charge inhomogeneities: Motivation and Approach}

It is a universally observed phenomenon in all hole-doped cuprate
systems that as the doping in the CuO\sb{2} planes is increased, the
in-plane Cu-O bonds shorten~\cite{radae;prb94i}.  This is seen in
diffraction measurements as a reduction in the $a$ and $b$ lattice
parameters with increasing doping.  This has a simple explanation from
the point of view that the energy band at the Fermi-level is a Cu
$3d_{x^2-y^2}$ - O $2p_{xy}$ $\sigma^*$ anti-bonding 
band~\cite{goode;f92}.  On doping
holes into this band, electron density is being removed from the
antibonding states which stabilizes the Cu-O bond resulting in a
shorter bond.  For example, in La$_{2-x}$Sr$_{x}$CuO$_{4}$ the
in-plane Cu-O bond shortens from 1.904~\AA\ to 1.882~\AA\ as $x$
changes from 0 to 0.2~\cite{radae;prb94i} and the average copper
valence changes from $2+$ to $\sim 2.2+$.

This observation has a profound implication if the charges are
inhomogeneously distributed in the CuO\sb{2} planes; for example, in
the presence of charge-stripes.  In this scenario, regions of the
plane are heavily doped (the stripes) and other regions are undoped
(the regions between the charge-rich stripes).    Thus,
{\it if the charges are inhomogeneously distributed} and {\it if the
charges are fluctuating slower than the lattice} so the lattice can
respond to the charge fluctuations, this implies that {\it there will
be a distribution of longer and shorter Cu-O bonds coexisting in the
structure.}  The same argument
holds whether or not the charge inhomogeneities are striped.
A high resolution measurement of the in-plane Cu-O bond
length {\it distribution} as a function of doping and temperature will
therefore reveal the existence, or otherwise, of charge
inhomogeneities.

Here we report high real-space resolution measurements of the 
in-plane Cu-O bond-length distribution as a function of doping and 
temperature in a series of compounds in the ``214'' family of cuprates:
La$_{2-x}$A$_x$CuO$_4$ (A=Sr,Ba).  We show that the distribution of
Cu-O bond lengths changes in a non-monotonic way with doping consistent
with the presence of charge inhomogeneities in the underdoped and
optimally doped material turning over to a homogeneous charge distribution
in the overdoped regime.  We also present supporting evidence from the
intermediate range structure which is completely consistent with there being
a microscopic coexistence of heavily doped and undoped regions of the CuO\sb{2}
plane in the underdoped, but superconducting, 214 materials.
This provides compelling {\it structural} evidence for the existence of 
microscopic charge inhomogeneities in these materials in the underdoped state.

\section{Experimental}
Samples of La$_{2-x}$Sr$_{x}$CuO$_{4}$ with $x=0.0$, 0.05, 0.10,
0.125, 0.15, 0.20, 0.25, and 0.30 were made using standard solid state
synthesis.  A sample with Ba replacing Sr with composition $x=0.15$
was also studied.  Sample preparation details are reported
elsewhere~\cite{bozin;prb99}.  Finally, a sample of
La$_{1.475}$Nd\sb{0.4}Sr\sb{0.125}CuO$_4$, also made by standard
solid-state reaction techniques, was studied.  Neutron powder
diffraction data as a function of doping were collected at 10~K on the
Special Environment Powder Diffractometer (SEPD) at the Intense Pulsed
Neutron Source (IPNS) at Argonne National Laboratory. Temperature
dependent data were collected on the Nd doped sample from the Glasses,
Liquids and Amorphous Diffractometer (GLAD) at IPNS and SEPD and on
the Ba doped sample and the Sr doped $x=0.125$ sample using the High
Intensity Powder Diffractometer (HIPD) at the Manuel Lujan Neutron
Scattering Center (MLNSC) at Los Alamos National Laboratory.  The
resolutions and backgrounds of each of these instruments are quite
different but they all give qualitatively the same results for the
T-dependence of the width of the in-plane Cu-O bond distribution.  In
each case, approximately 10g of finely powdered sample was sealed in a
cylindrical vanadium tube with He exchange gas.  The samples were
cooled using a closed-cycle He refrigerator. The data were corrected
for experimental effects and normalized, using the PDFgetN
program~\cite{peter;jac00}, to obtain the total structure function
$S(Q)$.  The
PDF, $G(r)$, is obtained by a Fourier transform of the data according
to $G(r) = {2\over \pi}\int_0^{\infty} Q[S(Q)-1]\sin Qr\>dQ$.
Representative data and PDFs from these samples are shown in
Fig.~\ref{fig;egdata}.

We are interested in extracting the width of the distribution of
in-plane Cu-O bond-lengths.  This information is contained in the
width of the first PDF peak at $\sim 1.9$~\AA .  The peak width comes
from the thermal and zero-point motion of the atoms plus any
bond-length distribution originating from charge inhomogeneities.  For
data collected as a function of doping at constant temperature (10K)
we expect the bond-length distribution (PDF peak-width) to be
constant or to vary weakly but smoothly with doping if the lattice is
softening or hardening.  What we actually see are very large changes
in the width of the bond-length distribution which are
non-monotonically varying with doping.  This behavior is straightforwardly
explained in the context of charge inhomogeneities as we describe below.
Furthermore, we argue that it cannot be explained as originating from
other extrinsic effects such as the strain resulting from chemical doping
or structural fluctuations coming from nearby structural phase
transitions.  The temperature dependent data further supports this
point of view: the onset of the Cu-O bond-length distribution
broadening does not correlate in any way with structural phase
transitions in these systems.

\section{Results}

\begin{figure}[tb]
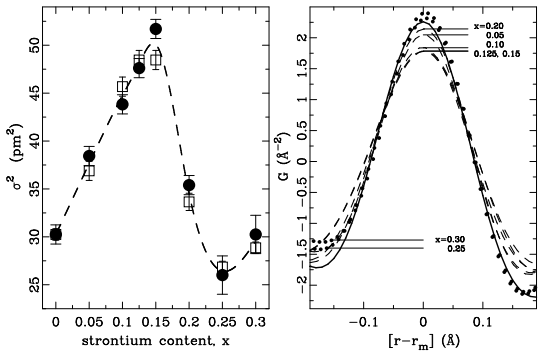

\begin{center}$\,$
\epsfxsize=1.4in
\epsfbox{PRLfig2NEW.fig}
\epsfxsize=1.4in
\epsfbox{PRLfig1NEW.fig}
\end{center}
\caption{(a) Solid circles: 
Peak width of the in-plane Cu-O PDF peak as a function of doping obtained
by fitting a Gaussian.
The data are plotted as
$\sigma^2$ where $\sigma$ is the Gaussian standard deviation. Open 
squares: inverse peak-height-squared scaled so
the $x=0.0$ points line up. Dashed line is a guide for the eye.
(b) PDF peak coming from the in-plane Cu-O bond for various doping
levels.  The peaks have been shifted 
so that their centers line up at $r_m = 1.91$~\AA .  
}
\protect\label{fig;widthvx}
\end{figure}

The squared width of the Cu-O bond-length 
distribution is plotted as a function of doping in Fig.~\ref{fig;widthvx}(a)
It is apparent in Fig.~\ref{fig;widthvx}(a) that the increase in the
mean-squared bond-length distribution is significant.  The
effect is large enough that it can be easily seen in the peaks
themselves.  In Fig.~\ref{fig;widthvx}(b) the in-plane Cu-O PDF peaks are
shown with their centroids lined up for convenient comparison of peak width.
The broadening is
readily apparent in this figure.

In Ref.~\cite{bozin;prl99;unpub} we argue in detail that the origin of this
behavior cannot be due to extrinsic effects but is readily explained
as originating from the presence of local charge inhomogeneities as would
be expected  in the presence of locally fluctuating charge stripes.  
We interpret the phenomenology in light of this model below.

The mean-square width of the Cu-O bond length distribution increases
monotonically (and almost linearly) with $x$ until $x=0.15$. Between
0.15 and 0.2 the peak abruptly sharpens and returns to the width of
the undoped sample by $x=0.25$ (Fig.~\ref{fig;widthvx}(a)).  If we
assume that in the structurally well-ordered, undoped, endmember
La\sb{2}CuO$_{4}$ there is a single, well-defined, in-plane Cu-O bond
length; then the width of the bond-length distribution due to quantum
zero-point motion is given by the width of the measured PDF peak
(after deconvoluting the experimental resolution function) for this
sample: $\sigma = 0.055$~\AA .  As the doping level is increased the
peak broadens smoothly as evident in Fig.~\ref{fig;widthvx}.  We
interpret this as originating from the presence of two distinct Cu-O
bonds: a shorter bond coming from more highly doped regions and a
longer bond from less highly doped (or undoped) regions of the
CuO\sb{2} plane.  This scenario is consistent with the stripe picture
for the cuprates but is also consistent with other forms of charge
inhomogeneity.  As doping is increased the broadening increases
because more short bonds appear.  The behavior changes between
$x=0.15$ and $x=0.20$ where the PDF peak is seen to abruptly sharpen.
At $x=0.20$ the peak width has almost completely returned to its
undoped (single bond length) value and at $x=0.25$ and $x=0.3$ it
remains sharp.  This behavior is interpreted as a crossover to a
regime where there is a homogeneous charge distribution in the
CuO\sb{2} planes and the electronic state is becoming more
Fermi-liquid like: the charge stripes (or inhomogeneities) have
disappeared by $x=0.2$.  Note that the data were collected at 10~K on
samples that are bulk superconductors; i.e., {\it the charge
inhomogeneities are observed in the superconducting state.}  It is
also interesting to note that the inhomogeneities and T\sb{c} are not
anti-correlated as they are for static stripes~\cite{ichik;cm99}: the
maximum T\sb{c} occurs approximately where the inhomogeneities are also
at a maximum.

\begin{figure}[tb]
\begin{center}$\,$
\epsfxsize=2.8in
\epsfbox{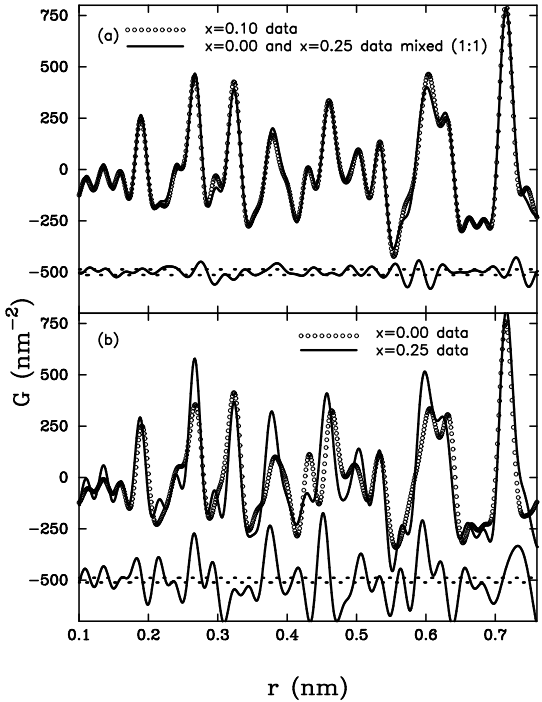}
\end{center}
\caption{(a) PDF from the $x=0.1$ data at 10~K (open
circles). The solid line shows the PDF obtained by making a linear
combination in a (1:1) ratio of the PDFs, shown in (b), from the
$x=0.0$ (open circles) and $x=0.25$ (solid line) samples.
The differences are plotted
below.  The dashed lines indicate expected uncertainties due to random
errors. }
\protect\label{fig;mix}
\end{figure}

If this picture is correct it has some implications for other features
in the intermediate range structure.  A coexistence of heavily doped
and undoped regions in the CuO\sb{2} planes implies a coexistence of
large and small CuO\sb{6} octahedral tilts: undoped
La$_{2-x}$Sr$_{x}$CuO$_{4}$ is heavily tilted (5$^\circ$ tilts) and
heavily doped La$_{2-x}$Sr$_{x}$CuO$_{4}$ is untilted.  The measured
PDF on the intermediate (nanometer) length-scale must be consistent
with the presence of tilt inhomogeneities if they exist.  We can make
a very simple test of this.  We have measurements of the PDFs of
heavily doped ($x=0.25$) and undoped ($x=0.0$) material.  Can we
explain the PDFs of an intermediate doped compound in the underdoped
region as a linear combination of these two PDFs?
Figure~\ref{fig;mix}(a)
shows a comparison of the PDF from the $x=0.1$ data-set with a linear
combination of the $x=0.00$ and the $x=0.25$ data-sets in a 1:1 ratio.
The $x=0.00$ and $x=0.25$ PDFs themselves are reproduced in
Fig.~\ref{fig;mix}(b) for comparison.  They are clearly very different
from each other (primarily because of the different CuO\sb{6}
octahedral tilt amplitudes~\cite{bozin;prb99}) yet when mixed they
reproduce the $x=0.1$ data-set very well.  Plotted below the PDFs are
difference curves.  The dotted lines above and below the difference
curves are the expected errors due to random counting statistics.  The
0.00/0.25 mixture reproduces the $x=0.1$ data set almost completely
within the expected uncertainties.  Clearly, the $x=0.1$ PDF is
consistent with the local environment in the CuO\sb{2} planes being a
mixture of heavily tilted and untilted octahedra.

\begin{figure}[tb]
\begin{center}$\,$
\epsfxsize=2.8in
\epsfbox{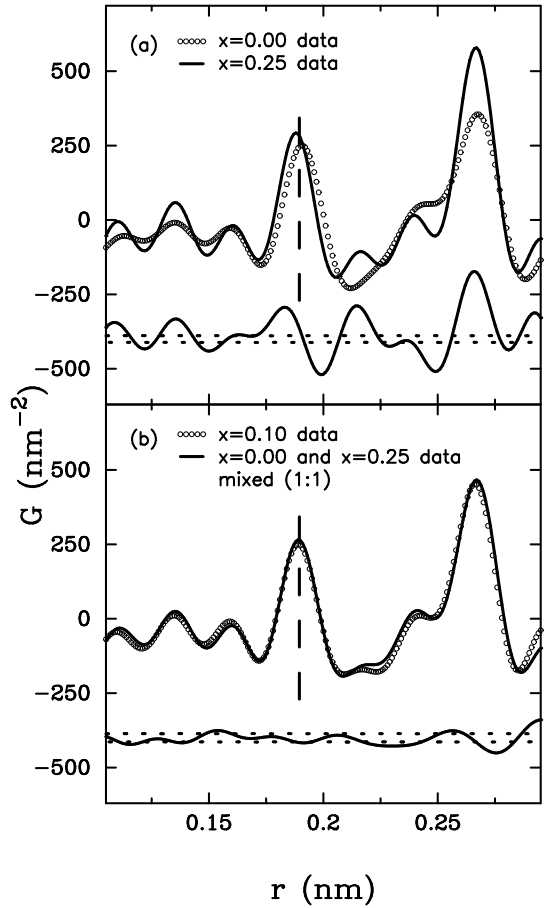}
\end{center}
\caption{(a) PDFs from the 10~K data from the $x=0.0$ (open circles) and 
$x=0.25$ (solid line) samples.
The difference is plotted below.  The dashed lines indicate expected 
uncertainties due to random errors. (b) PDF from $x=0.1$ data at 10~K (open
circles). The solid line shows the PDF obtained by making a linear combination
in a (1:1) ratio of the PDFs shown in (a). The vertical dashed lines show the
position of the peak centroid for the $x=0.1$ sample.}
\protect\label{fig;mixlor}
\end{figure}

It is also revealing to focus in on the in-plane Cu-O peak in the PDF.
This region of the PDF is shown on an expanded scale in Fig.~\ref{fig;mixlor}.
Reference to Fig.~\ref{fig;widthvx}(a) indicates that both the
$x=0.00$ and $x=0.25$ samples have narrow Cu-O nearest neighbor peaks
indicative of a single Cu-O bond-length broadened by zero point
motion.  The positions of these peaks are shifted with respect to each
other in these two compositions because of the shortening of the Cu-O
bond with doping.  This is evident in Fig.~\ref{fig;mixlor}(a) where
the PDFs from the $x=0.00$ and $0.25$ samples are shown: the peaks are
relatively sharp and their centroids are shifted.  In
Fig.~\ref{fig;mixlor}(b) it is clear that the broad peak centered on
an intermediate position in the $x=0.1$ sample is well reproduced as a
1:1 linear combination of the two sharp peaks shown in
Fig.~\ref{fig;mixlor}(a).  We note that the $x=0.00$ and $0.25$ PDFs
are not scaled or shifted at all when taking the linear combination.
Thus, both the position {\it and} broadening of the Cu-O bond-length
distribution of the $x=0.1$ sample, shown in
Fig.~\ref{fig;widthvx}(a), are self-consistently explained as a arising
from a mixture of heavily and undoped regions in the CuO\sb{2} plane.

\begin{figure}[tb]
\begin{center}$\,$
\epsfxsize=2.8in
\epsfbox{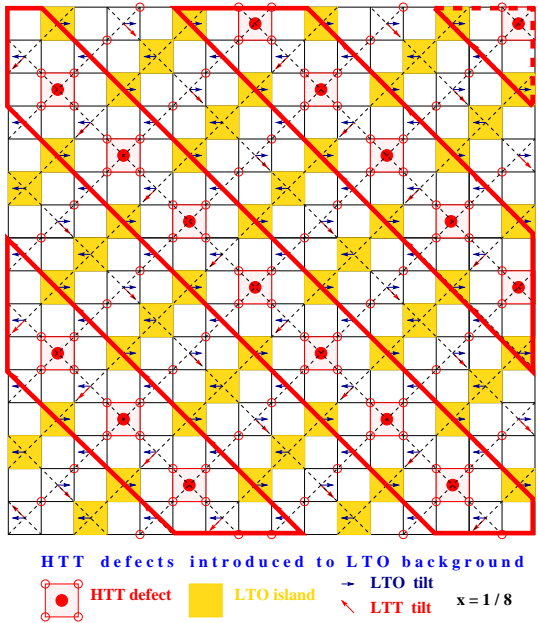}
\end{center}
\caption{Schematic representation of a topological model for
the CuO\sb{6} octahedral tilt directions in the
presence of charge stripes.  A plan view of the CuO\sb{2} plane is
shown.  Squares with dashed crosses are octahedra containing copper.
Localized holes are indicated by large circles and are arranged into
stripes.  Open circles on the shared vertices of the octahedra
indicate the in-plane oxygen atom at this vertex lies in the plane of
the CuO\sb{2} sheet.  Otherwise it is displaced up or down due to the
presence of octahedral tilts.  The underlying LTO symmetry tilt order
is indicated by the short arrows.  Longer arrows indicate the
direction of LTT-like tilts induced by the presence of the localized
charges which are presumed to create an untilted octahedron by locally
shortening the Cu-O bond length.  }
\protect\label{fig;topmod}
\end{figure}

There is one final piece of evidence supporting the presence of
octahedral tilt disorder in these samples.  Topological models of the
tilts of the corner-shared CuO$_6$ octahedra indicate that in addition
to tilt amplitude disorder there will be tilt-directional disorder
induced by localized holes~\cite{bozin;prb99}.  Regions of the plane
will be variously untilted, have [110] symmetry tilts (so-called
``LTO'' tilts), and have [100] symmetry (``LTT'') tilts.  This is
shown schematically in Fig.~\ref{fig;topmod}.
The topological modeling is intended to suggest how octahedral tilts
might reorient themselves in the presence of tilt defects.  Out-of-plane 
displacements of in-plane oxygen ions at the shared vertices of neighboring
octahedra are considered and the octahedra are assumed to be rigid.  A
predetermined set of tilt defects is introduced (in this case due to the
localized holes arranged in stripes, though this is clearly not the only
possibility) and the tilts of neighboring octahedra are determined by the new
boundary condition imposed through the shared oxygen due to the tilt of the
neighboring octahedron.  This approach clearly indicates how the presence of
untilted octahedra can introduce LTT-like tilts into the previously LTO tilted
background.

\begin{figure}[tb]
\begin{center}$\,$
\epsfxsize=2.8in
\epsfbox{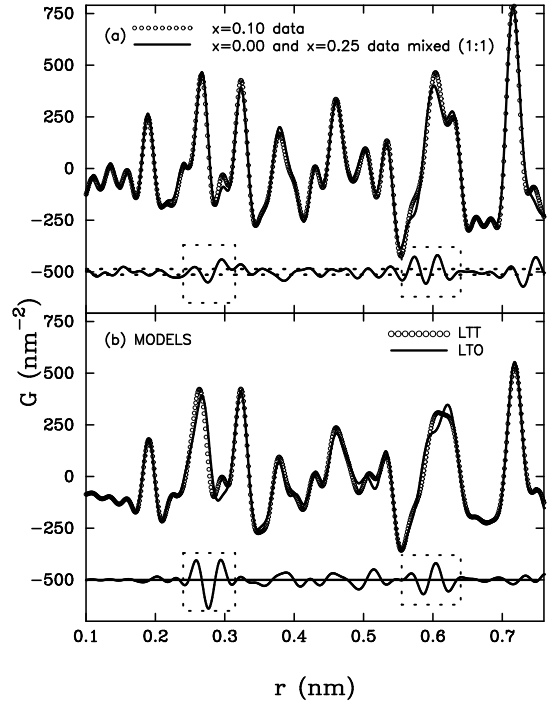}
\end{center}
\caption{(a) Comparison of the linear-combination PDF shown in 
Fig.~\protect\ref{fig;mix}(a) (solid line) with the $x=0.1$ data taken
at 10~K (open circles).
(b) Comparison of model PDFs calculated with
5$^\circ$ CuO$_6$ octahedral tilts which have LTT symmetry (open circles)
and LTO symmetry (solid line) respectively.  The difference curves are plotted
below the data.
}
\protect\label{fig;ltoltt}
\end{figure}

There is qualitative evidence in the PDF that
this tilt-directional disorder exists in these samples.  
Figure~\ref{fig;ltoltt}(a) again shows a comparison of the PDF obtained as 
a linear combination of $x=0.00$ and $x=0.25$ data with the PDF of the $x=0.1$
sample, this time plotted over a wider range of $r$.
The agreement is within the noise except in two regions around
$r=0.28$~nm and $r=0.59$~nm.  Fig~\ref{fig;ltoltt}(b) shows two model
PDFs calculated from models with 5$^\circ$ tilts of LTT symmetry (open
circles) and LTO symmetry (solid line), with the difference curve
plotted below.  The differences are fairly small, but significant
differences are observed around $r=0.28$~nm and $r=0.59$~nm which
coincides with the largest fluctuations seen in the difference curve
of Fig~\ref{fig;ltoltt}(a).  This is to be expected if some LTT
symmetry tilts are present in the $x=0.1$ data-set since the tilts in
the $x=0.0$ data-set are purely LTO
symmetry~\cite{radae;prb94i,bozin;prb99}.  This is highly suggestive,
supporting the idea that there is significant disorder with respect to
tilt amplitude {\it and} direction in underdoped
La$_{2-x}$Sr$_{x}$CuO$_{4}$, consistent with the presence of charge
inhomogeneities.

This observation presents a rather natural explanation for a puzzling
result which was reported from an earlier PDF study of
La$_{2-x}$Ba$_{x}$CuO$_{4}$~\cite{billi;prl94}.  In that
letter~\cite{billi;prl94} the local structure of the material was
studied as the sample went through the LTO-LTT structural phase
transition.  No change was observed in the local structure at this
transition, even though the average symmetry of the tilts was changing
from being tilted along [110] (LTO) to [100] (LTT) directions.  This
should involve a reorientation of local tilts which was not observed
in the data.  The long-range order of the tilts was changing at the
LTO-LTT structural transition but not the short-range order. It was
pointed out that the LTO symmetry could be obtained on average from a
linear combination of two different LTT-variants (e.g. [100]+[010])
and it was suggested that the LTO phase may be an inhomogeneous
mixture of differently oriented LTT domains.  However, it was not
clear at that time how, or why, the structure would prefer a
disordered tilt phase at low temperature.  The existence of
electronically driven charge inhomogeneities provides a rather natural
explanation of this observation. In light of the current results, we
believe that the picture of the LTO phase in the doped materials being
a mixture of two LTT variants is too simple.  However, a natural
explanation comes from the idea of the CuO\sb{2} planes being made up
of highly inhomogeneous octahedral tilts, including untilted regions
and heavily tilted regions with both LTO and LTT type tilts
present~\cite{bozin;prb99} and stabilized by the presence of charge
inhomogeneities; shown schematically in Fig.~\ref{fig;topmod}.

Currently we are investigating the temperature dependence of the local
structure to see whether the appearance of charge inhomogeneities can
be seen as a function of temperature. Preliminary results suggest that
there is an onset temperature where the inhomogeneities appear which
depends on doping.  In Nd-codoped La$_{2-x}$Sr$_{x}$CuO$_{4}$ this
temperature is close to that where long-range charge order is
observed~\cite{tranq;prb96}.  It is also a comparable temperature (in
the range 60-120~K) in La$_{2-x}$Sr$_{x}$CuO$_{4}$ and
La$_{2-x}$Ba$_{x}$CuO$_{4}$ which do not exhibit long-range charge
ordering. These results will be reported elsewhere.  This temperature
is comparable to the temperature scales where NQR wipeout
effects~\cite{hunt;prl99} and anomalies in XANES~\cite{lanza;jpcm99}
and transport measurements~\cite{ichik;cm99} are observed.

The result agrees
qualitatively with XAFS data of Bianconi {\it et
al.}~\cite{bianc;prl96} which indicates a broadening of the Cu-O bond
distribution at low temperature in similar compounds, though our
interpretation of the data is quite different from theirs.  A similar
XAFS study by  Niem\"oller~\cite{niemo;pc98} did not find significant evidence
for a broadening of the Cu-O correlation though the error bars are not
small.  These authors place an upper limit on
bond length distributions due to doping of 0.06~\AA .  Their result is 
therefore not inconsistent with our observation of a bond-length difference
of 0.02~\AA\ between the heavily and light doped regions of the CuO\sb{2}
plane.  

There is also an interesting overlap with the observations of
anomalous phonon softening in
YBa\sb{2}Cu\sb{3}O\sb{7-\delta}~\cite{petro;cm00,egami;js00;unpub}.
In these measurements the Cu-O half-breathing mode shows an unexpected
temperature dependence, apparently breaking into two branches at low
temperature.  Cu-O breathing and half-breathing modes are exactly
those that the nearest-neighbor peak in the PDF are the most sensitive
to (these modes give relative displacements of Cu and nearest neighbor
oxygen atoms in directions parallel to the bond).  Splitting of this
mode into two branches implies a doubling of the unit cell occurring at
low temperature which has been explained within a picture where
charge-stripes appear~\cite{petro;cm00,egami;js00;unpub} (the
canonical static stripes observed by Tranquada {\it et
al.}~\cite{tranq;n95} led to a unit cell quadrupling).  There is also
a softening of this half-breathing mode, in the sense that phonon
intensity moves to lower frequency at lower
temperature~\cite{petro;cm00,egami;js00;unpub}. Both these results
appear to be in qualitative agreement with the results obtained from
this PDF study in the La$_{2-x}$Sr$_x$CuO$_4$ system suggesting that
this behavior is somewhat universal.

\section{conclusions}
To summarize, we have presented local structural evidence from neutron
diffraction data which strongly supports the idea that doped charge in
the CuO$_2$ planes of superconducting La$_{2-x}$Sr$_{x}$CuO$_{4}$ for
$0<x \leq 0.15$ and at 10~K is inhomogeneous.  For doping levels of
$x=0.2$ and above the charge distribution in the Cu-O plane becomes
homogeneous.  This presumably reflects a crossover towards more
Fermi-liquid like behavior in the overdoped regime. The
inhomogeneities set in at low-temperatures comparable to the charge-ordering
temperature in Nd-codoped samples~\cite{tranq;prb96} and to the temperature
where NQR wipeout effects are observed related to spin 
freezing~\cite{hunt;prl99}.

\section{acknowledgments}

This work was supported financially by NSF through grant DMR-9700966 and
by the Sloan Foundation.
The experimental data were collected at the IPNS at Argonne National
Laboratory, which is funded by the US 
Department of Energy under Contract W-31-109-ENG-38, and at the 
MLNSC at Los Alamos National Laboratory which is funded by the department
of energy under contract W-7405-ENG-36.


\begin{thebibliography}{10}

\bibitem{tranq;prl94}
J.~M. Tranquada, D.~J. Buttrey, V.~Strachan, and J.~E. Lorenzo,
\newblock Phys. Rev. Lett. {\bf 73}, 1003 (1994).

\bibitem{sacha;prb95}
V.~Sachan, D.~J. Buttrey, J.~M. Tranquada, J.~E. Lorenzo, and G.~Shirane,
\newblock Phys. Rev. B {\bf 51}, 12742 (1995).

\bibitem{chen;prl96}
C.~H. Chen and S.-W. Cheong,
\newblock Phys. Rev. Lett. {\bf 76}, 4042 (1996).

\bibitem{mori;n98}
S.~Mori, C.~H. Chen, and S.-W. Cheong,
\newblock Nature {\bf 392}, 473 (1998).

\bibitem{tranq;n95}
J.~M. Tranquada, B.~J. Sternlieb, J.~D. Axe, Y.~Nakamura, and S.~Uchida,
\newblock Nature {\bf 375}, 561 (1995).

\bibitem{tranq;prb96}
J.~M. Tranquada, J.~D. Axe, N.~Ichikawa, Y.~Nakamura, S.~Uchida, and
  B.~Nachumi,
\newblock Phys. Rev. B {\bf 54}, 7489 (1996).

\bibitem{tranq;prl97i}
J.~M. Tranquada, J.~D. Axe, N.~Ichikawa, A.~R. Moodenbaugh, Y.~Nakamura, and
  S.~Uchida,
\newblock Phys. Rev. Lett. {\bf 78}, 338 (1997).

\bibitem{ichik;cm99}
N.~Ichikawa, S.~Uchida, J.~M. Tranquada, T.~Niem\"oller, P.~M. Gehring, {S.-H.
  Lee}, and J.~R. Schneider,
\newblock cond-mat/9910037.

\bibitem{zaane;prb89}
J.~Zaanen and O.~Gunnarson,
\newblock Phys. Rev. B {\bf 40}, 7391 (1989).

\bibitem{schul;prl89}
H.~J. Schulz,
\newblock Phys. Rev. Lett. {\bf 64}, 1445 (1989).

\bibitem{emery;pc93}
V.~J. Emery and S.~A. Kivelson,
\newblock Physica C {\bf 209}, 597 (1993).

\bibitem{white;prl98}
S.~R. White and D.~J. Scalapino,
\newblock Phys. Rev. Lett. {\bf 80}, 1272 (1998).

\bibitem{marti;cm00}
G.~M. Martins, C.~Gazza, J.~C. Xavier, A.~Feiguin, and E.~Dagotto,
\newblock cond-mat/0004316.

\bibitem{emery;pnas99}
V.~J. Emery, S.~A. Kivelson, and J.~M. Tranquada,
\newblock Proc. Natl. Acad. Sci. USA {\bf 96}, 8814 (1999).

\bibitem{bianc;ssc97}
A.~Bianconi, A.~Valletta, A.~Perali, and N.~L. Siani,
\newblock Solid State Commun. {\bf 102}, 369 (1997).

\bibitem{caste;cm00}
C.~Castellani, {C. Di Castro}, M.~Grilli, and A.~Perali,
\newblock cond-mat/0001231.

\bibitem{gorko;js99}
L.~Gor'kov,
\newblock J. Supercond. {\bf 12}, 9 (1999).

\bibitem{ovchi;prb99}
Y.~N. Ovchinnikov, S.~A. Wolf, and V.~Kresin,
\newblock Phys. Rev. B {\bf 60}, 4329 (1999).

\bibitem{hasse;prl99}
N.~Hasselmann, {A. H. Castro Neto}, C.~M. Smith, and Y.~Dimashko,
\newblock Phys. Rev. Lett. {\bf 82}, 2135 (1999).

\bibitem{goode;prb94}
J.~B. Goodenough and {J.-S. Zhou},
\newblock Phys. Rev. B {\bf 49}, 4251 (1994).

\bibitem{phill;pmb99}
J.~C. Phillips,
\newblock Philos. Mag. B {\bf 79}, 527 (1999).

\bibitem{marki;jpcs97}
R.~S. Markiewicz,
\newblock J. Phys. Chem. solids {\bf 58}, 1179 (1997).

\bibitem{mason;prl92}
T.~E. Mason, G.~Aeppli, and H.~Mook,
\newblock Phys. Rev. Lett. {\bf 68}, 1414 (1992).

\bibitem{thurs;prb92}
T.~R. Thurston, P.~M. Gehring, G.~Shirane, R.~J. Birgeneau, M.~A. Kastner,
  Y.~Endoh, M.~Matsuda, K.~Yamada, H.~Kojima, and I.~Tanaka,
\newblock Phys. Rev. B {\bf 46}, 9128 (1992).

\bibitem{yamad;prb98}
K.~Yamada, C.~H. Lee, K.~Kurahashi, J.~Wada, S.~Wakimoto, S.~Ueki, H.~Kimura,
  Y.~Endoh, S.~Hosoya, G.~Shirane, R. J.~Birgeneau, M.~Greven, M.~A. Kastner, and
  Y.~J. Kim,
\newblock Phys. Rev. B {\bf 57}, 6165 (1998).

\bibitem{mook;n00}
H.~A. Mook, P.~Dai, F.~Do\v{g}an, and R.~D. Hunt,
\newblock Nature {\bf 404}, 729 (2000).

\bibitem{hunt;prl99}
A.~W. Hunt, P.~M. Singer, K.~R. Thurber, and T.~Imai,
\newblock Phys. Rev. Lett. {\bf 82}, 4300 (1999).

\bibitem{crawf;prb91}
M.~K. Crawford, R.~L. Harlow, E.~M. {McCarron}, W.~E. Farneth, J.~D. Axe,
  H.~Chou, and Q.~Huang,
\newblock Phys. Rev. B {\bf 44}, 7749 (1991).

\bibitem{buchn;pc91}
B.~B\"uchner, M.~Braden, M.~Cramm, W.~Schlabitz, W.~Schnelle, O.~Hoffels,
  W.~Braunisch, R.~M\"uller, G.~Heger, and D.~Wohlleben,
\newblock Physica C {\bf 185-189}, 903 (1991).

\bibitem{egami;b;utbp00}
T.~Egami and S.~J.~L. Billinge,
\newblock {\em Underneath the Bragg Peaks},
\newblock Plenum, 2000,
\newblock to be published.

\bibitem{wagne;jncs78}
C.~N.~J. Wagner,
\newblock J. Non-Crystalline Solids {\bf 31}, 1 (1978).

\bibitem{billi;prb93}
S.~J.~L. Billinge and T.~Egami,
\newblock Phys. Rev. B {\bf 47}, 14386 (1993).

\bibitem{peter;jac00}
P.~F. Peterson, M.~Gutmann, T.~Proffen, and S.~J.~L. Billinge,
\newblock to be published in J. Appl. Crystallogr.  (2000).
\newblock See also,
  http://www.pa.msu.edu/cmp/billinge-group/programs/PDFgetN.

\bibitem{proff;prb99}
{Th. Proffen}, R.~G. DiFrancesco, S.~J.~L. Billinge, E.~L. Brosha, and G.~H.
  Kwei,
\newblock Phys. Rev. B {\bf 60}, 9973 (1999).

\bibitem{gutma;prb99}
M.~Gutmann, S.~J.~L. Billinge, E.~L. Brosha, and G.~H. Kwei,
\newblock Phys. Rev. B {\bf 61}, 11762 (2000).

\bibitem{petko;prl99}
V.~Petkov, {I-K. Jeong}, J.~S. Chung, M.~F. Thorpe, S.~Kycia, and S.~J.~L.
  Billinge,
\newblock Phys. Rev. Lett. {\bf 83}, 4089 (1999).

\bibitem{mikke;prl82}
J.~C. Mikkelson and J.~B. Boyce,
\newblock Phys. Rev. Lett. {\bf 49}, 1412 (1982).

\bibitem{radae;prb94i}
P.~G. Radaelli, D.~G. Hinks, A.~W. Mitchell, B.~A. Hunter, J.~L. Wagner,
  B.~Dabrowski, K.~G. Vandervoort, H.~K. Viswanathan, and J.~D. Jorgensen,
\newblock Phys. Rev. B {\bf 49}, 4163 (1994).

\bibitem{goode;f92}
J.~B. Goodenough,
\newblock Ferroelectrics {\bf 130}, 77 (1992).

\bibitem{bozin;prb99}
E.~S. Bo{\v z}in, S.~J.~L. Billinge, G.~H. Kwei, and H.~Takagi,
\newblock Phys. Rev. B {\bf 59}, 4445 (1999).

\bibitem{bozin;prl99;unpub}
E.~S.~Bo{\v z}in, S.~J.~L. Billinge, H.~Takagi, and G.~H. Kwei,
\newblock Phys. Rev. Lett.  (2000),
\newblock To be published.

\bibitem{billi;prl94}
S.~J.~L. Billinge, G.~H. Kwei, and H.~Takagi,
\newblock Phys. Rev. Lett. {\bf 72}, 2282 (1994).

\bibitem{lanza;jpcm99}
A.~Lanzara, G.~M. Zhao, N.~L. Saini, A.~Bianconi, K.~Conder, H.~Keller, and
  K.~A. M\"{u}ller,
\newblock J. Phys: Condens. Matter {\bf 11}, L541 (1999).

\bibitem{bianc;prl96}
A.~Bianconi, N.~L. Saini, A.~Lanzara, M.~Missori, T.~Rossetti, H.~Oyanagi,
  H.~Yamaguchi, K.~Oka, and T.~Ito,
\newblock Phys. Rev. Lett. {\bf 76}, 3412 (1996).

\bibitem{niemo;pc98}
T.~Niem{\" o}ller, B.~B{\" u}chner, M.~Cramm, C.~Huhnt, L.~Tr{\" o}ger, and
  M.~Tischer,
\newblock Physica C {\bf 299}, 191 (1998).

\bibitem{petro;cm00}
Y.~Petrov, T.~Egami, R.~J. McQueeney, M.~Yethiraj, H.~A. Mook, and F.~D{\v
  o}gan,
\newblock cond-mat/0003414.

\bibitem{egami;js00;unpub}
T.~Egami, Y.~Petrov, and D.~Louca,
\newblock this volume.

\end{thebibliography}


\end{document}